\documentstyle{article}
\begin{document}

\overfullrule=0pt
\baselineskip=18pt

\title{The Effect of Evanescent Modes and Chaos on Deterministic
Scattering in an Electron Waveguide.}
\author{G.B. Akguc and L.E. Reichl\\
Center for Studies in Statistical Mechanics and Complex Systems\\
The University of Texas at Austin\\
Austin, Texas 78712}

\date{\today}

\maketitle
\begin{abstract}

Statistical properties of Wigner delay times and the effect of
evanescent modes on the deterministic scattering of an electron
matter wave from a
classically chaotic 2-d electron waveguide are studied
for the case of
2, 6, and 16 propagating modes. Deterministic reaction matrix
theory
for this system is generalized to include the effect of evanescent modes on
the
scattering process.  The
statistical properties of the Wigner delay times for the deterministic
scattering process  are compared to
the predictions of random reaction matrix theory.

\end{abstract}


\bigskip
\bigskip

\section{Introduction}

In the 1950's it was observed that  nuclear scattering processes can have
statistical properties
indistinguishable from random scattering processes \cite{kn:wig}.  The
first hint that these random elements in the nuclear scattering
data might be due to underlying chaos in the nuclear dynamics
appeared in a paper by McDonald and Kauffman~\cite{kn:kauf} who studied
the energy level statistics for closed
   quantum billiards whose classical counterparts are either integrable or
chaotic. They found
that the quantized energy levels of the chaotic billiard  had a
statistical distribution which matched predictions of random matrix
theory.  The
first studies of the scattering properties of completely chaotic  quantum
systems with few degrees of freedom were due to Smilansky \cite{kn:blu}
   and since then a number of
papers have appeared \cite{kn:dor},\cite{kn:bar} analysing quantum
scattering using semi-classical
techniques \cite{kn:lin},\cite{kn:gas},\cite{kn:rou}, and focused on the
semiclassical regime. Recently Akguc and Reichl
\cite{kn:akg}  studied deterministic quantum scattering from a
chaotic  billiard,  in a regime where only a few channels are
open, using finite elements techniques and  have found random signatures
in the Wigner  delay times.

The analysis of fully quantum mechanical scattering processes, in systems
where only a few channels are open, is not easily accessible because this
regime is numerically demanding.
This fact has lead to renewed interest in the reaction matrix
formulation of scattering theory that was developed by Wigner
and Eisenbud
\cite{kn:2} in the late 1940's \cite{kn:3}. The idea behind reaction matrix
theory is to decompose  configuration space into a reaction region (cavity)
and an asymptotic scattering region (lead).
The exact wavefunction in the reaction region can be expanded in terms of any
convenient complete set of states with fixed  boundary conditions on the
surface of the reaction region, provided  the coupling between the
reaction region (cavity) and asymptotic scattering region is
singular  \cite{kn:4}, \cite{kn:5}.  Reaction matrix theory provides
a convenient
framework for predicting the scattering properties of systems governed by
random Hamiltonian matrices. We shall call the theory that uses reaction
matrix theory to
predict the scattering properties of systems with Gaussian random
Hamiltonians, {\it random reaction matrix theory} or RRMT. The predictions of
RRMT have been compared to experimental nuclear scattering
data \cite{kn:por},
scattering in electron waveguides \cite{kn:mar},
and resonances in acoustic and
micro-wave resonators \cite{kn:sto}, under conditions in which these
systems are
thought to have classically chaotic dynamics. These predictions, in
turn, can be
compared to the scattering properties of chaotic systems.
RRMT, as it is currently formulated, neglects some possibly important
effects in
the scattering process, namely the
effect of evanescent modes and some of the energy dependence of
resonance poles.

In this paper, we will study the deterministic scattering of an electron in a
two dimensional electron waveguide, which has a classically chaotic cavity
formed by a ripple billiard connected to a lead at one end (see Figure
1). This type of cavity is particularly well suited to the use of
reaction matrix theory, because a simple coordinate transformation
allows us to construct the basis states inside the cavity by
diagonalizing a Hamiltonian matrix.  We will generalize the reaction
matrix theory
for two dimensional waveguides to
include the
effect of evanescent modes. The effect of evanescent modes on
scattering processes has
been studied for nuclear scattering processes
\cite{kn:wig2},\cite{kn:mey},\cite{kn:tre},\cite{kn:ing} using
approximate
theories. For electron waveguides, we can include these effects
exactly.  We will show
that for the waveguide we consider, evanescent modes dominate the scattering
properties of the waveguide in energy regions where new propagating
channels open. We will  also compare the statistical properties of
Wigner delay times for the
deterministic waveguide scattering process to the predictions of RRMT.

We begin, in Section 2, by developing the reaction matrix theory of
deterministic scattering in our electron waveguide, starting from a
configuration
space formulation rather than the usual eigenmode formulation, and we
construct the
Hamiltonians for the cavity (reaction region) and leads (asymptotic scattering
region) of an electron waveguide. In Section (3) (and Appendix A) we use the
hermiticity of the total Hamiltonian to compute the strength of the coupling
between the cavity and lead. In Section (4), we derive the reaction
matrix. In
Section (5), we derive the scattering matrix. In Section (6), we describe the
method we use to obtain a complete set of basis states for a cavity
with a rippled
wall. In Section (7) we discuss the accuracy of the reaction matrix
predictions
by comparing them with a finite element calculation. In Section (8), we
discuss
the effect of evanescent modesß on the scattering process, and in
Section (9) we compare the statistical properties of the Wigner delay
times for deterministic scattering in the waveguide with predictions of
RRMT.  Finally, in Section (10),
we make some concluding remarks.

\bigskip
\bigskip
\section{Scattering Hamiltonian}

We will consider the scattering properties of an electron with mass,
$m$,  in the
waveguide shown in Fig. (1). The electron enters from the left with
energy E along an infinitely long straight lead
which has infinitely hard walls. The electron wave is reflected back
to the left  by an
infinitely hard
wall located at $x=0$. The scattering is strongly affected by the region
$0<x<L$ (the cavity)
in which the upper wall is rippled.

The Schrodinger equation, which describes propagation of a particle wave,
$\Psi(x,y,t)$, for all times, $t$, is given by
\begin{equation}
i{\hbar}{{\partial}{\Psi}(x,y,t)\over {\partial}t}=
{\biggl[}-{{\hbar}^2\over 2m}{\biggl(}{{\partial}^2\over
{\partial}x^2}+{{\partial}^2\over
{\partial}y^2}{\biggr)}+V(x,y){\biggr]}{\Psi}(x,y,t),
\end{equation}
where ${\hbar}$ is Planck's constant. The potential, $V(x,y)$, has the
following
properties: $V(x,y)={\infty}$ for $(L{\leq}x<\infty)$;
$V(x,0)=\infty$ for $(-\infty{\leq}x{\leq}L)$;
$V(x,y=g(x))=\infty$ for $(0<x<L)$; and
$V(x,y=a+d)=\infty$ for $(-\infty<x<0)$;
where $g(x)=d+a{\cos}(4{\pi}x/L)$ gives the contour of the ripple, $d$ is the
average width of the cavity, $L$ is the length, and $a$ is the ripple
amplitude. Throughout this paper, we take the electron mass to be the
effective mass
of an electron in GaAs, $m=0.067m_e$, where $m_e$ is the free electron mass.

We can introduce projection operators, ${\hat
P}={\int_{-\infty}^{0}}dx~{\int_{-\infty}^{\infty}}dy~~|x,y{\rangle}
{\langle}x,y|$ and ${\hat
Q}={\int_{0}^{L}}dx~{\int_{-\infty}^{\infty}}dy~~|x,y{\rangle}{\langle
}x,y|$ which satisfy the completeness relation
   ${\hat
Q}+{\hat P}={\hat 1}$ (all wavefunctions are zero for $L<x$). Here
$|x,y{\rangle}$
is the simultaneous eigenstate of position operators, ${\hat x}$ and
${\hat y}$.
The projection operators, ${\hat Q}$  and ${\hat P}$
have the property
that ${\hat Q}={\hat Q}^2$, ${\hat P}={\hat P}^2$, and ${\hat Q}{\hat
P}={\hat P}{\hat Q}=0$.
If a state,
$|{\Psi}{\rangle}$ has spatial dependence,
${\Psi}(x,y){\equiv}{\langle}x,y|{\Psi}{\rangle}$, over the interval
$(-\infty<x<L)$,
then the state ${\langle}x,y|{\hat Q}|{\Psi}{\rangle}={\Psi}(x,y)$ for
$(0<x<L)$ and the
state ${\langle}x,y|{\hat P}|{\Psi}{\rangle}={\Psi}(x,y)$ for
$(-\infty<x<0)$.

Inside the cavity (Region I, $(0<x<L)$ in Fig.(1)), we define a Hamiltonian,
\begin{equation}
{\hat H}_{QQ}{\equiv}{\hat Q}{\biggl[}
{1\over 2m}({\hat p}_x^2+{\hat p}_y^2)+V(x,y){\biggr]}{\hat Q},
\end{equation}
where ${\hat p}_x$ and ${\hat p}_y$ are
momentum operators and
$m$ is the mass of the particle. The Hamiltonian, ${\hat H}_{QQ}$, is
Hermitian
and therefore it will have a complete, orthonormal set of eigenstates which we
denote as
${\hat Q}|{\phi}_j{\rangle}$.  We can write the eigenvalue equation in the
region, $0<x<L$, as
${\hat H}_{QQ}{\hat Q}|{\phi}_j{\rangle}={\lambda}_j{\hat
Q}|{\phi}_j{\rangle}$, where
${\lambda}_j$ is the $j^{th}$ energy eigenvalue of ${\hat H}_{QQ}$ and
$j=1,2,...M$ (we will later let $M{\rightarrow}{\infty}$).
Because there is an infinitely hard wall at $x=L$, the eigenstates
${\phi}_j(x,y){\equiv}{\langle}x,y|{\hat Q}|{\phi}_j{\rangle}$ must be zero at
$x=L$. We have some freedom
in choosing the boundary condition
at $x=0$. In this paper, we will require that the eigenstates,
${\phi}_j(x,y)$,
have zero slope at $x=0$ so that ${d{\phi}_j\over dx}{\bigr|}_{x=0}=0$.
Singular coupling, between the cavity and the lead, will correct for
the fact that
the actual wavefunction does not have zero slope at
$x=0$. The completeness of the states,
${\hat Q}|{\phi}_j{\rangle}$, allows us to
write the completeness relation,
${\sum_j}{\hat Q}|{\phi}_j{\rangle}{\langle}{\phi}_j|{\hat Q}={\hat Q}$.
Orthonormality requires
that ${\langle}{\phi}_j|{\hat Q}|{\phi}_{j'}{\rangle}={\delta}_{j,j'}$.
The part, inside the cavity, of any state,
$|{\Psi}{\rangle}$, in the waveguide
can be expanded in terms of this complete set of states, so that
\begin{equation}
{\hat Q}|{\Psi}{\rangle}={\sum_{j=1}^M}
{\langle}{\phi}_j|{\hat Q}|{\Psi}{\rangle}~|{\phi}_j{\rangle}.
\label{eq:interior1}
\end{equation}

Inside the lead (Region II, ($-\infty<x<0$) in Fig. (1)), we define a
Hamiltonian
\begin{equation}
{\hat H}_{PP}{\equiv}{\hat P}{\biggl[}
{1\over 2m}({\hat p}_x^2+{\hat p}_y^2)+V(x,y){\biggr]}{\hat P}.
\end{equation}
Its eigenvalues are
continuous and have range, $(0{\leq}E{\leq}\infty)$. The eigenvector of
${\hat H}_{PP}$, with eigenvalue, $E$, will be denoted ${\hat
P}|E{\rangle}$. The eigenvalue equation then reads,
${\hat H}_{PP}{\hat P}|E{\rangle}=E{\hat P}|E{\rangle}$. Because the leads
are assumed to be straight, the transverse parts of the energy eigenstates,
in the
leads, decouple from the longitudinal part.  Because the walls of the
channels are
infinitely hard, the energy eigenstates in the leads (for $x<0$) can be
written
\begin{equation}
{\hat P}|E{\rangle}={\sum_{n=0}^{N}}{\Gamma}_n{\hat P}|{\Phi}_n{\rangle},
\label{eq:exterior1}
\end{equation}
where ${\Gamma}_n={\langle}{\Phi}_n|{\hat P}|E{\rangle}$ and
\begin{equation}
{\langle}x,y|{\hat P}|{\Phi}_n{\rangle}={\chi}_{k,n}(x)\sqrt{2\over d+a}
{\sin}{\biggl(}{n{\pi}y\over d+a}{\biggr)} \label{eq:exterior2}
\end{equation}
represents the contribution to ${\langle}x|E{\rangle}$ from the
$n^{th}$ transverse quantum state in the lead. Although we have summed over
the first $N$
transverse states we will later let $N{\rightarrow}\infty$. For a particle
with energy, $E$,
the state,
${\hat P}|{\Phi}_n{\rangle}$, has the property that
\begin{equation}
{\hat H}_{PP}{\hat P}|{\Phi}_n{\rangle}=E{\hat P}|{\Phi}_n{\rangle}=
{{\hbar}^2\over 2m}{\biggl(}k_n^2+
{\biggl(}{n{\pi}\over d+a}{\biggr)}^2{\biggr)}{\hat P}|{\Phi}_n{\rangle}.
\label{eq:exterior3}
\end{equation}
where
\begin{equation}
E={{\hbar}^2\over 2m}{\bigl(}k_n^2+
{\biggl(}{n{\pi}\over d+a}{\bigr)}^2{\biggr)}. \label{eq:exterior4}
\end{equation}
The state, ${\hat P}|{\Phi}_n{\rangle}$ is called the $n^{th}$ channel.
Eq. (\ref{eq:exterior4}) gives  the decomposition of the total energy,
    $E$, into its longitudinal and
transverse parts when the electron is in the channel, ${\hat
P}|{\Phi}_n{\rangle}$.
For a given energy, $E$, there are an infinite number of channels for
the particle, some propagating and some evanescent. Channels with
propagating modes
occur if the longitudinal wavevector is real. Channels with
evanescent modes occur if
the longitudinal wavevector is pure imaginary.  Evanescent modes describe
localized
contributions to the electron states in the waveguide.  There are
an infinite
number of them and for some values of the energy, $E$, they play a dominant
role in
determining the dynamics in the waveguide \cite{Boese}.

We couple the cavity and the lead at their interface, $x=0$,
via the singular operator, ${\hat V}=C{\delta}({\hat x}){\hat p}_x$. The
coupling constant, $C$, can be determined by the condition that the total
Hamiltonian \cite{kn:8} be Hermitian (see Section (3)).
Then
\begin{equation}
{\hat H}_{QP}={\hat
Q}{\hat V}{\hat
P}=C{\hbar\over
i}{\int_0^L}dx_1{\int_{-\infty}^{\infty}}dy_1~{\int_{-\infty}^{0}}dx_0
~|x_1,y_1{\rangle}
     {\delta}(x_1-x_0)
     {\delta}(x_0){d\over dx_0}{\langle}x_0,y_1|, \label{eq:surf1}
\end{equation}
and
\begin{equation}
{\hat H}_{PQ}={\hat
P}{\hat V}{\hat
Q}=C{\hbar\over
i}{\int_{-\infty}^{0}}dx_0{\int_{-\infty}^{\infty}}dy_0~{\int_0^L}dx_1
~|x_0,y_0{\rangle}
{\delta}(x_0-x_1)
     {\delta}(x_1){d\over dx_1}{\langle}x_1,y_0|. \label{eq:surf2}
\end{equation}
It is useful to remember that ${\int_0^L}dx~{\delta}(x)={1\over 2}$,
${\int_0^L}dx~{\delta}(x-x_0)=1$ if $0<x_0<L$, and
${\int_0^L}dx~{\delta}(x-x_0)=0$ if $L<x_0$ or $x_0<0$.
Note also that
\begin{equation}
{\hat H}_{QQ}={\int_0^L}dx~{\int_{-\infty}^{\infty}}dy~|x,y{\rangle}
{-{\hbar}^2\over 2m}{\biggl(}{d^2\over dx^2}+{d^2\over dy^2}+V(x,y){\biggr)}
{\langle}x,y|, \label{eq:surf3}
\end{equation}
and
\begin{equation}
{\hat
H}_{PP}={\int_{-\infty}^{0}}dx~{\int_{-\infty}^{\infty}}dy~|x,y{\rangle}
{-{\hbar}^2\over 2m}{\biggl(}{d^2\over dx^2}+{d^2\over dy^2}+V(x,y){\biggr)}
{\langle}x,y| \label{eq:surf4}
\end{equation}

The total Hamiltonian of the system can be written
\begin{equation}
{\hat H}={\hat H}_{QQ}+{\hat H}_{PP}+{\hat H}_{QP}+{\hat H}_{PQ}.
\end{equation}
The waveguide energy eigenstates, $|E{\rangle}$, satisfy the eigenvalue
equation
${\hat H}|E{\rangle}=E|E{\rangle}$. The states, $|E{\rangle}$, can be
decomposed into
their contributions
from the two regions of configuration space, so that
\begin{equation}
|E{\rangle}={\sum_{j=1}^M}{\gamma}_j{\hat Q}|{\phi}_j{\rangle}
+{\sum_{n=0}^{N}}{\Gamma}_n{\hat P}|{\Phi}_n{\rangle}, \label{eq:eneig1}
\end{equation}
where ${\gamma}_j={\langle}{\phi}_j|{\hat Q}|E{\rangle}$ and
${\Gamma}_n={\langle}{\Phi}_n|{\hat P}|E{\rangle}$. The eigenvalue
equation then
takes the form
\begin{equation}
\pmatrix{{\hat H}_{QQ}&0&\ldots&{\hat H}_{QP}&{\hat H}_{QP}&\ldots\cr
0&{\hat H}_{QQ}&\ldots&{\hat H}_{QP}&{\hat H}_{QP}&\ldots\cr
\vdots&\vdots&\ddots&\vdots&\vdots&   \cr
{\hat H}_{PQ}&{\hat H}_{PQ}&\ldots&{\hat H}_{PP}&0&\ldots\cr
{\hat H}_{PQ}&{\hat H}_{PQ}&\ldots&0&{\hat H}_{PP}&\ldots\cr
\vdots&\vdots&\ldots&\vdots&\vdots&\ldots\cr}
\pmatrix{{\gamma}_1{\hat Q}|{\phi}_1{\rangle}\cr
{\gamma}_2{\hat Q}|{\phi}_2{\rangle}\cr
\vdots\cr
{\Gamma}_1{\hat P}|{\Phi}_1{\rangle}\cr
{\Gamma}_2{\hat P}|{\Phi}_2{\rangle}\cr
\vdots\cr}
=E\pmatrix{{\gamma}_1{\hat Q}|{\phi}_1{\rangle}\cr
{\gamma}_2{\hat Q}|{\phi}_2{\rangle}\cr
\vdots\cr
{\Gamma}_1{\hat P}|{\Phi}_1{\rangle}\cr
{\Gamma}_2{\hat P}|{\Phi}_2{\rangle}\cr
\vdots\cr}. \label{eq:eneig2}
\end{equation}
This yields a series of equations
\begin{equation}
{\hat H}_{QQ}{\hat Q}|{\phi}_j{\rangle}{\gamma}_j+{\sum_{n=0}^{N}}{\hat
H}_{QP}
{\hat P}|{\Phi}_n{\rangle}{\Gamma}_n=
E{\hat Q}|{\phi}_j{\rangle}{\gamma}_j, \label{eq:eneig3}
\end{equation}
for $j=1,2,...,M$ and
\begin{equation}
{\hat H}_{PP}{\hat P}|{\Phi}_n{\rangle}{\Gamma}_n
+{\sum_j}H_{PQ}{\hat Q}|{\phi}_j{\rangle}{\gamma}_j
=E{\hat P}|{\Phi}_n{\rangle}{\Gamma}_n \label{eq:eneig4}
\end{equation}
for $n=1,2,...,N$.
The  condition for Hermiticity of
the Hamiltonian, ${\langle}{\Psi}_{\beta}|{\hat H}|{\Psi}_{\alpha}{\rangle}
={\langle}{\Psi}_{\alpha}|{\hat H}|{\Psi}_{\beta}{\rangle}^*$,
    allows us to determine that the value of the coupling
constant, $C$, is $C={4{\hbar}i\over 2m}$ (see Appendix A).

\section{The Reaction Matrix}

We now have enough information to derive the reaction matrix for this system.
Let us first multiply  Eq. (\ref{eq:eneig3}) by ${\langle}{\phi}_j|{\hat
Q}$ to obtain
\begin{equation}
{\langle}{\phi}_j|{\hat H}_{QQ}|{\phi}_j{\rangle}{\gamma}_j+
{\sum_{n=1}^N}{\langle}{\phi}_j|{\hat H}_{QP}|{\Phi}_n{\rangle}{\Gamma}_n=
E{\langle}{\phi}_j|{\hat Q}|{\phi}_j{\rangle}{\gamma}_j
\end{equation}
which reduces to
\begin{equation}
({\lambda}_j-E){\gamma}_j+C{{\hbar}\over 4i}
{\sum_{n=1}^N}{\phi}_{j,n}^*(0){d{\chi}_n\over dx}{\biggr|}_0{\Gamma}_n=0.
\label{eq:reac1}
\end{equation}
If we use Eq. (\ref{eq:coupconst}) for the coupling constant, $C$, we can
rewrite
Eq. (\ref{eq:reac1}) and obtain the following expression for ${\gamma}_j$,
\begin{equation}
{\gamma}_j={{\hbar}^2\over 2m}{1\over (E-{\lambda}_j)}{\sum_{n=1}^N}
{\phi}_{j,n}^*(0){d{\chi}_n\over dx}{\biggr|}_a{\Gamma}_n. \label{eq:reac2}
\end{equation}
The continuity equation (\ref{eq:cont1}), when applied to energy
eigenstates, yields
\begin{equation}
{\Gamma}_n{\chi}_n(0)={\sum_{j=1}^M}{\gamma}_j{\phi}_{j,n}(0)=
{\sum_{n'=1}^N}~{R}_{n,n'}~{d{\chi}_{n'}\over dx}
{\biggr|}_0{\Gamma}_{n'}. \label{eq:reac3}
\end{equation}
where
\begin{equation}
{R}_{n,n'}={{\hbar}^2\over 2m}{\sum_{j=1}^M}
{{\phi}_{j,n'}^*(0){\phi}_{j,n}(0)\over (E-{\lambda}_j)}
\label{eq:reac4}
\end{equation}
is the $(n,n')^{th}$ matrix element of the reaction matrix.

We must now distinguish between propagating and evanescent modes. The
states in the leads, for
propagating modes, can be written
\begin{equation}
{\Gamma}_n{\chi}_n(x)=\frac{a_n}{\sqrt{k_n}}
{\rm e}^{-ik_nx}+\frac{b_n}{\sqrt{k_n}}{\rm e}^{ik_nx},
\label{eq:chan1}
\end{equation}
where
\begin{equation}
k_n=\sqrt{{2mE\over {\hbar}^2}-{\biggl(}{n{\pi}\over d+a}{\biggr)}^2}
\label{eq:chan2}
\end{equation}
If there are $\nu$ propagating modes then $n=1,2,...,\nu$. Here we use a unit
current normalization.
The  evanescent modes in the leads can be written
\begin{equation}
{\Gamma}_n{\chi}_n(x)=\frac{c_n}{\sqrt{k_n}}{\rm e}^{-{\kappa}_n|x|},
\label{eq:chan3}
\end{equation}
where
\begin{equation}
{\kappa}_n=\sqrt{{\biggl(}{n{\pi}\over d+a}{\biggr)}^2-{2mE\over
{\hbar}^2}} \label{eq:chan4}
\end{equation}
For evanescent modes the index $n={\nu}+1,{\nu}+2,...,N$ where
$N{\rightarrow}\infty$.
\section{The Scattering Matrix}

To obtain the scattering matrix, we must first separate the propagating
modes from the
evanescent modes. This first step is accomplished as follows.
Using Eq. (\ref{eq:chan1}) and Eq. (\ref{eq:chan3}) we can
write Eq. (\ref{eq:reac3}) in the matrix form
\begin{equation}
\pmatrix{{\bar a}+{\bar b}\cr  {\bar c}\cr}=
\pmatrix{{\bar K_p}&0\cr  0&{\bar K_e}\cr}
\pmatrix{{\bar R}_{pp}&{\bar R}_{pe}\cr  {\bar R}_{ep}&{\bar R}_{ee}\cr}
\pmatrix{{\bar K}_p&0\cr  0&{\bar K}_e\cr}
\pmatrix{i({\bar b}-{\bar a})\cr  {\bar c}\cr},
\label{eq:scat1}
\end{equation}
where

\begin{eqnarray}
{\bar a}=\pmatrix{a_1\cr
\vdots\cr
a_{\nu}\cr},~~~
{\bar b}=\pmatrix{b_1\cr
\vdots\cr
b_{\nu}\cr}, ~~~
{\bar c}=\pmatrix{c_{\nu +1}\cr
\vdots\cr
c_N\cr}, \nonumber\\
{\bar K}_p=\pmatrix{{\sqrt{k_1}}&\ldots&0\cr \vdots&\ddots&\vdots\cr
0&\ldots&{\sqrt{k_{\nu}}}\cr},~~~
{\bar K}_e=\pmatrix{{\sqrt{{\kappa}_1}}&\ldots&0\cr
\vdots&\ddots&\vdots\cr  0&\ldots&{\sqrt{{\kappa}_{N-\nu}}}\cr} \nonumber\\
{\bar R}_{pp}=\pmatrix{R_{1,1}&\ldots&R_{1,{\nu}}\cr
\vdots&\ldots&\vdots\cr
R_{{\nu},1}&\ldots&R_{{\nu},{\nu}}\cr},~~~
{\bar R}_{pe}=\pmatrix{R_{1,{\nu}+1}&\ldots&R_{1,N}\cr
\vdots&\ldots&\vdots\cr
R_{{\nu},{\nu}+1}&\ldots&
R_{{\nu},N}\cr}, \nonumber\\
{\bar R}_{ep}=\pmatrix{R_{{\nu}+1,1}&\ldots&R_{{\nu}+1,{\nu}}\cr
\vdots&\ldots&\vdots\cr
R_{N,1}&\ldots&R_{N,{\nu}}\cr},~~~
{\bar R}_{ee}=\pmatrix{R_{{\nu}+1,{\nu}+1}&\ldots&R_{{\nu}+1,N}\cr
\vdots&\ldots&\vdots\cr
R_{N,{\nu}+1}&\ldots&R_{N,N}\cr}
     \label{eq:scat2}
\end{eqnarray}
If we expand out Eq. (\ref{eq:scat1}), we find
\begin{eqnarray}
{\bar a}+{\bar b}=i{\bar K}_p{\bar R}_{pp}{\bar K}_p({\bar b}-{\bar a})
+{\bar K}_p{\bar R}_{pe}{\bar K}_e{\bar c} \label{eq:scat3}
\end{eqnarray}
\begin{eqnarray}
{\bar c}=i{\bar K}_e{\bar R}_{ep}{\bar K}_p({\bar b}-{\bar a})+
{\bar K}_e{\bar R}_{ee}{\bar K}_e{\bar c}.
\label{eq:scat4}
\end{eqnarray}
    From Eq. (\ref{eq:scat4}) we can write  ${\bar c}$  as
\begin{equation}
{\bar c}={i\over ({\bar 1}_e-{\bar K}_e{\bar R}_{ee}{\bar K}_e)}
{\bar K}_e{\bar R}_{ep}{\bar K}_p({\bar b}-{\bar a}), \label{eq:scat5}
\end{equation}
where ${\bar 1}_e$ is a unit matrix with the same dimensions as ${\bar
R}_{ee}$.
If we substitute Eq. (\ref{eq:scat5}) into Eq. (\ref{eq:scat3}), we find
\begin{equation}
{\bar a}+{\bar b}=i{\bar D}({\bar b}-{\bar a}), \label{eq:scat6}
\end{equation}
where
\begin{equation}
{\bar D}={\biggl[}{{\bar K}_p\bar R}_{pp}{\bar K}_p
+{\bar K}_p{\bar R}_{pe}{\bar K}_e{1\over ({\bar 1}_e-
{\bar K}_e{\bar R}_{ee}{\bar K}_e)}
{\bar K}_e{\bar R}_{ep}{\bar K}_p{\biggr]}. \label{eq:scat7}
\end{equation}
The second term on the right in Eq. (\ref{eq:scat7}) contains the effect of
the
evanescent states on the propagating modes in the waveguide. The {\it
scattering
matrix}, ${\bar S}$, relates the outgoing propagating modes to the incoming
propagating modes through the relation, ${\bar a}={\bar S}{\bar b}$.
The scattering
matrix is thus given
by
\begin{equation}
{\bar S}=-{({\bar 1}_p-i{\bar D})\over ({\bar 1}_p+i{\bar D})},
\label{eq:scat8}
\end{equation}
where ${\bar 1}_p$ is a unit matrix with the same dimension as ${\bar
R}_{pp}$.
We see from Eqs. (\ref{eq:scat7}) and (\ref{eq:scat8}), that the evanescent
modes may play an important role in the scattering process. To see this
effect
on the resonance structure of ${\bar S}$ matrix,
    we obtain a more explicit form  as follows.
First we define the coupling matrices,
\begin{equation}
{\bar w}_{Np}\equiv
\pmatrix{\phi_{11}& \ldots& \phi_{1p}\cr
\vdots&&\vdots \cr
\phi_{N1}& \ldots& \phi_{Np}\cr }{\bar K}_p ~~~{\rm and}~~~
{\bar w}_{Ne}\equiv
\pmatrix{\phi_{11}& \ldots& \phi_{1e}\cr
\vdots&&\vdots \cr
\phi_{N1}& \ldots& \phi_{Ne}\cr }{\bar K}_e,
\label{eq:scat9}
\end{equation}
where $p$ is the number of propagating modes and $e$ is
the number of evanescent modes in the lead.
The matrix ${\bar D}$ can be written in terms of the coupling matrices as,
\begin{eqnarray}
{\bar D}&=&{\bar w}_{pN}^\dagger \frac{1}{E{\bar 1}_N-{\bar H}_{in}}
{\bar w}_{Np}
\nonumber
\\ +&{\bar w}_{pN}^\dagger& \frac{1}{E{\bar 1}_N-{\bar H}_{in}}
{\bar w}_{Ne}\frac{1}{{\bar 1}_e-{\bar w}_{eN}^\dagger\frac{1}{E{\bar 1}_N
-{\bar H}_{in}}{\bar w}_{Ne}}{\bar w}_{eN}^\dagger \frac{1}{E{\bar 1}_N-{\bar
H}_{in}}{\bar w}_{Np}
\label{eq:scat10}
\end{eqnarray}
where ${\bar H}_{in}$ is a diagonal matrix formed by the eigenvalues of
the Hamiltonian, $H_{QQ}$, inside the cavity. The second part of Eq.
(\ref{eq:scat10}) can
be  rearranged by expanding in a series and regrouping terms,
\begin{eqnarray*}
{\bar D}&=&{\bar w}_{pN}^\dagger \frac{1}{E{\bar 1}_N-{\bar H}_{in}}
{\bar w}_{Np}
\nonumber
\\ +&{\bar w}_{pN}^\dagger& \frac{1}{E{\bar 1}_N-{\bar H}_{in}}
{\bar w}_{Ne}{\bar w}_{eN}^\dagger \frac{1}{E{\bar
1}_N-{\bar H}_{in}-{\bar w}_{Ne}{\bar w}_{eN}^\dagger}{\bar w}_{Np}.
\label{eq:scat10.5}
\end{eqnarray*}
This expression for ${\bar D}$ can then be substituted into Eq.
(\ref{eq:scat8}) and
we obtain the following form for the scattering matrix,
\begin{eqnarray}
{\bar S}=-(1-2i{\bar w}_{pN}^\dagger \frac{1}{E{\bar 1}_N-{\bar
H}_{in}-{\bar w}_{Ne}{\bar w}_{eN}^\dagger +i{\bar w}_{Np}{\bar
w}_{pN}^\dagger}{\bar
w}_{Np})
\label{eq:scat11}
\end{eqnarray}
As  can be seen from the denominator of this expression, evanescent
modes affect the positions of resonance poles in complex energy plane
because of
their dependence on both coupling matrices, ${\bar w}_{pN}$ and
${\bar w}_{eN}$.
Contributions
from term,
${\bar w}_{Ne}{\bar w}_{eN}^\dagger $ are not included in RRMT
calculations. Also, the  energy dependence of the coupling matrix,
${\bar w}_{pN}$ is  neglected in RRMT calculations, although this is
known to various authors and they simply assume that the energy regions they
consider are far from channel openings. As
we see in Eq. (\ref{eq:scat11}), evanescent modes may play an
    important  role in the scattering process, and in
    subsequent sections, we
will investigate their effect  on scattering of an electron from the ripple
cavity.

\section{Basis States for the Cavity Region}

We now describe a method to obtain the complete set of eigenstates, ${\hat
Q}|\phi_j{\rangle}$, of the Hamiltonian, $H_{QQ}$. We will require that these
states have zero slope at the at the cavity-lead interface ($x=0$).
We introduce a coordinate transformation which straightens the rippled wall of
the cavity \cite{kn:akg}, \cite{kn:lun}. Then we can obtain a Hamiltonian
matrix which
can be diagonalized to find the eigenvalues, ${\lambda}_j$, and eigenstates,
${\phi}_j(x,y)$.
The first step is to write the eigenvalue equation, $H_{QQ}{\hat
Q}|\phi_j{\rangle}={\lambda}_j{\hat Q}|\phi_j{\rangle}$ in configuration
space.
It takes the form
\begin{equation}
{-{\hbar}^2\over 2m}{\biggl(}{d^2\over dx^2}+{d^2\over
dy^2}+V(x,y){\biggr)}{\phi}_j(x,y)={\lambda}_j{\phi}_j(x,y),
\end{equation}
where ${\phi}_j(x,y){\equiv}{\langle}x,y|{\hat
Q}|\phi_j{\rangle}$.
After the coordinate change,
\begin{equation}
u=x, \;\;\;\;\;\; v=\frac{y}{d+a~ \cos(\frac{4\pi}{L}  x)},
\end{equation}
     we obtain an eigenvalue equation in
terms of the coordinates, $u$ and $v$, given by
\begin{equation}\bar{H}\psi_j(u,v) \equiv
-\frac{\hbar^2}{2m}(\partial_u^2+h_1\partial_v^2+h_2\partial_{uv}^2+
h_3\partial_v)\psi_j(u,v)={\lambda}_j\psi_j(u,v)
\label{eq:hb}
\end{equation}
where
\begin {eqnarray*}
h_1 = \frac{1 + v^2 g_u^2}{g^2},
h_2 = \frac{-2 v g_u}{g} ,
h_3 = \frac{-v g_{uu}}{g} + \frac{2 v g_u^2}{g^2},
\end{eqnarray*}
$g=g(u)\equiv d + a \cdot \cos(\frac{4\pi}{L} u)$,
$g_u \equiv \frac{\partial g}{\partial u}$, and
${\psi}_j(u,v)={\phi}_j(x(u,v),y(u,v))$.  The boundary conditions in $(u , v)$
space are given by
$\partial_u\psi_l(0,v)=0$, $\psi_l(L,v)=0$,
$\psi_l(u,0)=0$, and
$\psi_l(u,1)=0$, so that in terms of these coordinates the walls are straight.
Note that in the $(u,v)$ coordinate frame, the states, ${\psi}_j(u,v)$ are
normalized with a weighting factor, $g(u)$, so that
\begin{equation}
\int\!\!\int
     g(u)\,\psi_j^\dagger(u,v)\psi_{j'}(u,v)\,du\,dv=\delta_{j,j'},
\end{equation}

The state, $\psi_j(u,v)$, can be expanded in terms of a Fourier basis,
\begin{equation}
\psi_j(u,v)=\sum_{m=1}^{\infty} \sum_{n=1}^{\infty} B^j_{mn} \phi_{mn}(u,v)
\end{equation}
with
\begin{equation}
\phi_{mn}(u,v)=\frac{2}{\sqrt{L}} g^{-1/2} \sin(n \pi v)
\cos(\frac{(2 m-1)\pi u}{2 L}), \nonumber
\end{equation}
where $B^j_{mn}$ are the unknown expansion coefficients.
As a result of this expansion, the boundary value problem is
transformed into the
eigenvalue problem,
\begin{equation}
\sum_{m=1}^{\infty} \sum_{n=1}^{\infty} H_{mnm'n'} B^j_{m'n'}=
E_j B^j_{mn}.
\end{equation}

The Hamiltonian matrix elements, $H_{mnm'n'}$, are given by
\begin{equation}
H_{mnm'n'}=\frac{4}{L}
\int_0^L\!du \int_0^1\! dv \sqrt{g} ~ sin(n \pi v)~ f ~
\bar{H}~ (sin(n' \pi v) f'/\sqrt{g})
\label{eq:cof}
\end{equation}
where $f \equiv \cos(\frac{(2 m-1) \pi u}{2 L})$,
$f'\equiv \cos(\frac{(2 m'-1)\pi u}{2 L})$,
$g\equiv d+ a~ \cos(\frac{4\pi}{L}  u)$, and $\bar{H}$ is the
differential operator defined
in Eq. ~(\ref{eq:hb}). Note that we cannot use
integration by parts to get a symmetrical form, as was done in Refs.
~\cite{kn:akg} and
~\cite{kn:lun}  because surface terms  will not drop out.

We calculate the Hamiltonian matrix elements using Eq.~(\ref{eq:cof}).
We reduce the double integral to a single integral after
integrating in the $v$ direction. After some algebra we
find the  following form which is suitable for numerical calculations,
\begin{eqnarray}
H_{mnm'n'} =
-w_{mm'}^1\delta_{nn'}/2+ (n \pi)^2 w_{mm'}^2\delta_{nn'}/2 +
(2 (n \pi)^2 -3)w_{mm'}^3\delta_{nn'} \nonumber \\
+w_{mm'}^4\delta_{nn'}/4
+(1-\delta_{nn'})(-1)^{n+n'}(\frac{4 w_{mm'}^3 n'^3 n}{(n^2-n'^2)^2}+
\frac{w_{mm'}^4 n n'}{(n^2-n'^2)})
\end{eqnarray}

where
\begin{eqnarray*}
w_{mm'}^1&\equiv&\int_0^L du
\frac{4 f_{uu} g^2 -4 f_u g_u g -2 f g_{uu} g +3 f g_u^2}{4 g^2} f'
\\
w_{mm'}^2&\equiv&\int_0^L du \frac{f}{g^2} f' ,
w_{mm'}^3\equiv\int_0^L du \frac{f g_u^2}{g^2}f'
\\
w_{mm'}^4&\equiv&\int_0^L du
\frac{-2 f_u g_u g + f g_u^2 - f g_{uu} g + 2 f g_u^2}{g^2} f'
\end{eqnarray*}
The eigenvalues and eigenvectors of ${\bar H}$ can be calculated efficiently
due to the sinusoidal integrals.
Eigenvectors of ${\bar H}$ give values for the expansion coefficients,
$B^j_{mn}$, and the
eigenfunctions in  u-v space can be found from these coefficients.
The solution can then be transformed back to x-y
space to obtain the basis states, ${\phi}_j(x,y)$.

\section{Accuracy of Reaction Matrix Theory Computations}

We have computed the scattering matrix and the amplitude of the evanescent
modes for the
waveguide  given by Fig. 1, using both the reaction matrix theory
presented in the previous sections, and an independent finite
element method (FEM) as a check on the reaction matrix results. For this
comparison we used the following parameter values:  $a=10\AA$, $d=100\AA$,
$L=200\AA$, and
$m=0.067m_e$ where
    $m_e$ is the free electron mass.
    This gives $E_n=n^2{{\hbar}^2{\pi}^2\over
    2m(d+a)^2}=(0.0622n^2)eV$ for the energy at which the $n^{th}$
    propagating mode appears.

We will discuss  our results for the energy interval,
$E_1<E<9E_1$, in which one propagating mode ($E_1<E<4E_1$) and two
propagating modes
($4E_1<E<9E_1$) can exist in
the leads. We studied both the case in which $a=0$, so the cavity
is rectangular, and  the case $a=10\AA$ in which the dynamics in the cavity is
fully chaotic \cite{kn:akg}.  We have computed
the partial Wigner delay times, ${\tau}_n={\hbar}{\frac {d\theta_n} {dE}}$,
where ${\theta}_n$ is the
$n^{th}$ eigenphase of the S-matrix.

For a rectangular cavity, $a=0$ (the upper boundaries of the rectangular
region is shown with a dotted line
in Fig.1), we can find analytic expressions for the S-matrix which
 serve as a
check on the accuracy of our programs.  For the energy regime where
only one mode
can propagate in the leads, the S-matrix (S-function in this case) is a
complex
number with unit magnitude and is given by the value of reflection
coefficient,
$S={\rm e}^{i2kL}$. This is the phase shift of the wave as it enters
the cavity and
reflects back to the entrance. The eigenphase of the S-matrix is
${\theta}=2kL$.  Since the phase angle depends linearly on $k$, no resonance
occurs. The Wigner delay time has no peaks.

The reaction matrix for a rectangular cavity can be written exactly. Since
there is no mode coupling in the rectangular cavity, it is enough to
calculate
$R_{11}$ in the energy regime where only one mode propagates. The
eigenvalues of
$H_{QQ}$ take the form ${{\hbar}^2\over 2m}{\bigl(}\frac{(m\cdot
\pi)^2}{d^2}+ \frac{((2\cdot n - 1)\cdot \pi)^2}{(2\cdot
L)^2}{\bigr)}$, where $m$
and $n$ positive integers representing the transverse and longitudinal
degrees of freedom. We obtain
\begin{eqnarray}
\label{eq:rexact}
R_{11} &=& \lim_{N \rightarrow \infty} \sum_{n=0}^{N} \frac{1}{e-(\pi/d)^2-
((2\cdot n -1)\cdot \pi/(2 L))^2}  \nonumber \\
    &=& \frac{\tan(L\cdot \sqrt(e-(\pi/d)^2))}{ \sqrt(e-(\pi/d)^2))}
=-\frac{\tan(k L)}{k},
\end{eqnarray}
where $e=2mE/{\hbar}^2$.
The S function is given in terms of $R_{11}$,
\begin{equation}
S=-\frac{1-i k R_{11}}{1+ i k R_{11}}=-\frac{1+i \tan(k L)}{1-i \tan(k
L)}={\rm e}^{i2kL},
\end {equation}
so the eigenphase is given by ${\theta}=2kL$ which coincides with the
exact result.

We can now check the accuracy of the methods we are using. In Fig. 2,
the sum in Eq.
(\ref{eq:rexact}) is truncated at $N=1000$ and $N=10000$ and used to compute
the reflection coefficient. The relative error is less than
$10^{-4}$ for the number of terms we kept in  following calculations.
The phase angle curve calculated by FEM is in agreement to the order
of $10^{-5}$
with the exact result. More discussion about the FEM method can be found in
ref
\cite{kn:akg} and references cited there.

Let us now consider the case of the ripple cavity. In Fig. 3, we show
the Wigner
delay times  calculated by using Eq. (\ref{eq:scat8}) and we compare them
to the Wigner delay times obtained from the FEM calculation. In Fig. 3(a)
we show the Wigner delay time, ${\tau}_1$,  in the energy interval where
there is one propagating mode in the leads. In Fig. 3(b), we  show two
partial Wigner delay times, $\tau_1$ and $\tau_2$, in the energy interval
where there are 2 modes in the lead. We have kept up to 2500 terms in Eq.
(\ref{eq:reac4}). We checked the accuracy of these eigenvalues  by
increasing the size of the Hamiltonian matrix and comparing
 eigenvalues of a
matrix with 5500 eigenvalues, and a matrix with 10,000 eigenvalues. We found
that the first 2500 eigenvalues were the same to an accuracy greater than
$10^{-5}$. We used the first 2500 eigenvalues and their eigenstates to
construct the reaction matrix.

We also looked at the analytic continuation of the S-matrix into the
lower complex energy plane to see the resonances explicitly.  In Fig. 4 we
used
 Eq. (\ref{eq:scat8}) (including evanescent
modes) to obtain the
analytic continuation of the S-matrix. The position of the three peaks, shown
in Fig. (4),
are $E_1=1.3585-i0.123$, $E_2=1.8991-i0.2403$, and $E_3=3.1241-i0.2316$.
The real part of peak position shows the resonance energy and the
imaginary part  shows its lifetime.  These are both consistent with Fig.
(3.a). The full width
at half maximum of resonances in Fig. (3.a) is in agreement with
imaginary parts
of  S-Matrix poles in Fig. (4). We obtained the following
numerical values for the resonance positions, $E_i$, and their widths,
${\Gamma}_i$ in Fig. (3.a), ($E_1=1.3733$,
$\Gamma_1=0.16$), ($E_2=1.8917$:
$\Gamma_2=0.20$), and  ($E_3=3.1359$, $\Gamma_3=0.21$), where
$\Gamma$ is defined
as the full width at half maximum. In the upper half plane one gets
corresponding
zeros of the S-Matrix.

\section{The Effect of Evanescent Modes}

For the shape of waveguide cavity that we consider here, the effect of
evanescent modes on Wigner delay times appears to be most important just
before a new channel opens in the lead. We have studied the effect of
evanescent modes using  parameters,
$a=10\AA$,
$d=100\AA$,
$L=500\AA$, and $m=0.067m_e$, and we use the form of the S-matrix in Eq.
(\ref{eq:scat8}) to obtain our results. We compare the variation of
S-matrix elements, $S_{ij}$, for an S-matrix which includes the evanescent
modes (${\bar K}_e{\neq}0$), with an S-matrix, ${\bar S}^0$, which excludes
evanescent modes (${\bar K}_e=0$). We use the cavity length, $L=500\AA$,
(rather than the samller length $L=200\AA$ used to check accuracy)
to increase the density of resonances in any given energy interval.
 In a mesh
based numerical method (like FEM or a finite difference) increasing
cavity length is numerically is not efficient due to the increasing number
of nodal points, but the reaction matrix approach can easily accomodate
longer cavitites. In Fig. (5.a) we show the effect of the evanescent modes
on the Wigner delay time at energies just below where the second channel
opens and, as we expect, there is a considerable increase in the delay of
the electron. The absolute value of the amplitude,
$c_1$, of the first evanescent mode also increases just before the second
propagating channel opens as shown in  Fig. (5.b).  The effect of the first
evanescent mode, $c_1$, is dominant since the amplitudes of the
second and higher
evanescent modes are near zero. A similar behavior of the evanescent modes
occurs at energies just below where the third propagating channel
opens. There again, one  evanescent mode
becomes dominant. In Figs. (5.c) and Fig. (5.d) we compare the behavior of
first and second partial Wigner delay times, ${\tau}_1$ and
${\tau}_2$, respectively, both for the case when the contributions of the
evanescent modes are included and  for the case  when they are
removed in calculations of the S-matrix in this energy regime.

In Fig.~6 we plot $|S_{11}|$ and $|S_{12}|$ in the energy interval,
$4E_1<E<9E_1$ (two propagating modes).
We find that $|S_{22}|=|S_{11}|$  and $|S_{21}|=|S_{12}|$.
Therefore we show only these two matrix elements.
In Fig. 7, we show the effect of evanescent
 modes on  S-matrix elements in the two mode energy regime by plotting the
differences, $|S_{11}^0|-|S_{11}|$ and $|S_{12}^0|-|S_{12}|$. The
difference in the magnitude of the S-matrix elements is small, but the
difference in the slopes can be fairly large.

We have also looked at the analytic continuation of S-matrix elements in
the complex energy plane and we find good agreement with the
predictions of Wigner delay time plots. In Fig.~8, we show partial Wigner
delay times in energy interval, $4E_1<E<9E_1$. In Fig.~9 we show the
behavior of $|S_{11}|$ in the complex energy plane.  Fig. 9.a gives large
scale behavior, and Fig. 9.b focuses on behavior near the real axis. The
poles near to real energy axis (shown in Fig. 9.b) determine the sharp
peaks in the Wigner delay times. The poles further from the real axis
determine the broader peaks in the Wigner delay time plots.

In Fig.~10 and in Fig.~11
 we show the effect of the energy dependence of the coupling
matrices,
${\bar w}_{Np}$ and ${\bar w}_{Ne}$. This energy dependence is always
neglected in RRMT calculations.  In Fig.~10, we plot $|S_{11}|$, both for
the case when the energy dependence of ${\bar w}_{Np}$ and ${\bar w}_{Ne}$
is taken into account (full line), and for the case when the energy
dependence of
${\bar w}_{Np}$ and ${\bar w}_{Ne}$ is fix at value, $E=6.5E_1$
(dotted-dashed line).
 In Fig.~11, the effect of the energy dependence of ${\bar w}_{Np}$ and
${\bar w}_{Ne}$ on the distributions of poles in the complex energy plane
is shown. The position of S-matrix poles changes when the variation with
energy of the coupling constants is not included. In Fig. 11, the solid
lines are contour lines of
$|S_{11}|$ for the reaction matrix calculation with evanescent modes
included and the energy dependence of
${\bar w}_{Np}$ and
${\bar
w}_{Ne}$ included as in Fig.~(6).  The dotted-dashed lines shows the same
quantity
but using  coupling matrices, ${\bar w}_{Np}$ and ${\bar w}_{Ne}$, with
dependence
on energy, $E$ fixed at the real value $E=6.5E_1$. Neglect of the energy
dependence of the coupling constants causes a shift of the poles away from
their true positions. This shift  is small in the neighborhood of the
fixed energy, $E=6.5E_1$, but it grows as one moves further away in energy.

\section{The Signatures of Chaos}
In this section we compute the statistical properties of the Wigner delay
times
obtained for deterministic scattering  of the electron
from the ripple cavity.  We consider only configurations of the
ripple cavity for
which the dynamics of the cavity is classically chaotic.
One can use either Eq.
(\ref{eq:scat8}) or Eq. (\ref{eq:scat11}) to calculate Wigner delay times
deterministically. We have checked that they give identical answers.  For the
deterministic calculations, we can get sufficient data to develop good
statistics
by changing the ripple size from
$a=10\AA$ to $a=30\AA$ in units of $0.2\AA$. In Figs. (9a), (9.b),
and (9.c), we
show the statistics for the total Wigner delay times for deterministic
scattering
for the cases when
$M=2$, $M=6$, and $M=16$ propagating modes, respectively, exist in the
leads. In
these Figures,
 $P(\tau)$ is the histogram of Wigner delay times normalized so the area is
equal to one,  and $<\tau>$ is the mean Wigner delay time. The distribution,
$P(\tau)$,  shifts from a Poisson-like distribution to
Gaussian-like distribution as  we increase the number of channels. For a small
number of channels the distribution is asymmetric and has a long tail.

We also looked at the statistics of the  total Wigner delay times  obtained by
replacing the S-matrix in Eq. (\ref{eq:scat11}), by the
equation
\begin{eqnarray}
{\bar S}=-(1+2ig^2{\bar w}^\dagger \frac{1}{E{\bar 1}_N-{\bar
H}_{in}^{'}-ig^2{\bar w}{\bar w}^\dagger}{\bar w}),
\end{eqnarray}
where ${\bar H}_{in}^{'}$ is chosen from an
Gaussian Orthogonal Ensemble (GOE) and the coupling matrix, ${\bar w}$, is
constructed
from M  eigenvectors of the realizations of ${\bar H}'_{in}$ in the same GOE
ensemble. Note that $g$ is a coupling constant that
must be determined from experiment. We
also checked our result by building ${\bar w}$ using the M eigenvectors of
each realization of ${\bar H}_{in}$ and we get a similar distribution for
the  corresponding number of channels.

We have calculated the Wigner delay time by taking the derivative of the
S-matrix eigenphase curve,
${\theta}(E)$ versus $E$, in two different ways. The first way is to
take two neighboring
energy points (we chose E=0 and E=0.001) and used these obtain one Wigner
delay time for each realization of $H_{in}^{'}$. The second way is to
obtain a whole seeries of Wigner delay times from the ${\theta}(E)$ versus
$E$ curve for a single  realization of
$H_{in}^{'} $. We have checked that these two methods give similar results
as we would expect due to ergodicity.

In Fig.~(12) we show the distribution of total Wigner delay times
obtained from the Gaussian Orthogonal Ensemble as described above.
The middle column with Figures Figs. (12.d), (12.e), and (12.f)
 corresponds to a
coupling constant, $g=1.8$, which is the strong coupling regime for
RRMT.  The right-most column with Figures Figs. (12.g), (12.h), and (12.i)
corresponds to a coupling constant
$g=1.0$.  The distribution of total Wigner delay times for our deterministic
scattering from the chaotic ripple cavity, agrees qualitatively with the
predictions of random matrix theory for strong coupling.  This is consistent
with the fact that the opening between the ripple cavity and the leads for our
case is very large.

\section {Conclusion}

In this paper we have studied the effect that evanescent modes have on the
scattering properties of an electron in a waveguide with a ``chaotic"
cavity. We
have reformulated the reaction matrix theory of electron waveguide
scattering to
explicitly include the effect of evanescent modes. We have found that
evanescent
modes can increase the delay of the electron for energies near the opening
of new
channels. This effect has been seen before \cite{Boese}.  The
scattering
system we have considered is relatively ``soft". There are no impurities
and no
sharp corners to snag evanescent modes, and yet their effect is still
noticeable.
For systems with impurities and sharp corners, we expect the effect of
evanescent
modes to be even more dramatic.

We have also studied the effect of neglecting the energy dependence of the
coupling matrices that appear in the reaction matrix approach to
scattering. This
appears to cause an efective repulsion on the positions of quasibound state
poles.

The effects of both the evanescent modes and the energy dependence of coupling
matrices are routinely neglected in RRMT, and this should be kept in mind
when attempting to use that theory to make predictions about real waveguide
scattering
experiments or numerical simulation of deterministic waveguide scattering
systems.

We have also studied the statistical distribution of the Wigner delay times
for
scattering from our chaotic waveguide cavity, for the case of $M=2$, $M=6$ and
$M=16$ propagating modes. To build adequate statistics for comparison with
RRMT
predictions, we have included data for a range of ripple amplitudes, being
careful to include data only from the regime where the internal dynamics of
the
ripple cavity is completely chaotic. If the ripple amplitude is too large
or too
small, the cavity will again develop a mixed phase space \cite{kn:lun}. We
find
fairly good qualitative agreement with the predictions of strong coupling
RRMT.

\section {Acknowledgements}

The authors wish to thank the Welch Foundation, Grant
No.F-1051, NSF Grant INT-9602971  and DOE contract No.DE-FG03-94ER14405
for partial support of this work. We also thank the University of
Texas at Austin High Performance Computing Center for use of their computer
facilities, and we thank German Luna-Acosta and Thomas Seligman for helpful
comments.

\pagebreak

\pagebreak

\newpage

\section{Appendix A: Hermiticity Condition}

Consider the arbitrary states, $|{\Psi}_{\alpha}{\rangle}$ and
$|{\Psi}_{\beta}{\rangle}$.
The condition for Hermiticity of these states is that
\begin{equation}
{\langle}{\Psi}_{\beta}|{\hat H}|{\Psi}_{\alpha}{\rangle}
-{\langle}{\Psi}_{\alpha}|{\hat H}|{\Psi}_{\beta}{\rangle}^*=0.
\end{equation}
We will use this condition to determine the coupling constant $C$
(this method of determining the strength of the coupling was first
suggested by
Pavlov ~\cite{kn:8}). We can decompose the states
$|{\Psi}_{\alpha}{\rangle}$ and
$|{\Psi}_{\beta}{\rangle}$
into their contributions to the two disjoint configuration space
regions and write them
in the form,
\begin{equation}
|{\Psi}_{\alpha}{\rangle}={\hat Q}|{\Psi}_{\alpha}{\rangle}+{\hat
P}|{\Psi}_{\alpha}{\rangle}={\sum_{j=1}^M}{a}_j{\hat
Q}|{\phi}_j{\rangle} +{\sum_{n=1}^{N}}A_n{\hat P}|{\Phi}_n{\rangle},
\end{equation}
where ${a}_j={\langle}{\phi}_j|{\hat Q}|{\Psi}_{\alpha}{\rangle}$ and
$A_n={\langle}{\Phi}_n|
{\hat P}|{\Psi}_{\alpha}{\rangle}$,
and
\begin{equation}
|{\Psi}_{\beta}{\rangle}={\hat Q}|{\Psi}_{\beta}{\rangle}+{\hat
P}|{\Psi}_{\beta}{\rangle}={\sum_{j=1}^M}{b}_j{\hat
Q}|{\phi}_j{\rangle} +{\sum_{n=1}^{N}}B_n{\hat P}|{\Phi}_n{\rangle},
\end{equation}
where ${b}_j={\langle}{\phi}_j|{\hat Q}|{\Psi}_{\beta}{\rangle}$ and
$B_n={\langle}{\Phi}_n|
{\hat P}|{\Psi}_{\beta}{\rangle}$,
     Inside the cavity, $0{\leq}x<L$, we have
expanded $|{\Psi}_{\alpha}{\rangle}$ and $|{\Psi}_{\beta}{\rangle}$ in
terms of the complete set of energy
eigenstates, ${\hat Q}|{\phi}_j{\rangle} $,  of the Hamiltonian, ${\hat
H}_{QQ}$.
In the lead, $-\infty{\leq}x<0$, we have
expanded $|{\Psi}_{\alpha}{\rangle}$ and $|{\Psi}_{\beta}{\rangle}$ in
terms of the complete set of energy
eigenstates, ${\hat P}|{\Phi}_{n}{\rangle}$,  of the Hamiltonian, ${\hat
H}_{PP}$.

The Hermiticity condition takes the form
\begin{eqnarray}
{\langle}{\Psi}_{\beta}|{\hat H}|{\Psi}_{\alpha}{\rangle}
-{\langle}{\Psi}_{\alpha}|{\hat H}|{\Psi}_{\beta}{\rangle}^* \nonumber\\
={\sum_{j=1}^M}[{\langle}{\phi}_j|{\hat H}_{QQ}|{\phi}_j{\rangle}
-{\langle}{\phi}_j|{\hat H}_{QQ}|{\phi}_j{\rangle}^*]a_jb_j^* \nonumber\\
{\sum_{j=1}^M}{\sum_{n=1}^N}[{\langle}{\phi}_j|{\hat H}_{QP}|{\Phi}_n{\rangle}
-{\langle}{\Phi}_n|{\hat H}_{PQ}|{\phi}_j{\rangle}^*]a_j^*B_n \nonumber\\
+{\sum_{j=1}^M}{\sum_{n=1}^N}[{\langle}{\Phi}_n|{\hat
H}_{PQ}|{\phi}_j{\rangle}
-{\langle}{\phi}_j|{\hat H}_{QP}|{\Phi}_n{\rangle}^*]b_jA_n^* \nonumber\\
+{\sum_{n=1}^N}[{\langle}{\Phi}_n|{\hat H}_{PP}|{\Phi}_n{\rangle}
-{\langle}{\Phi}_n|{\hat H}_{PP}|{\Phi}_n{\rangle}^*]A_n^*B_n=0.
\label{eq:herm2}
     \end{eqnarray}

We can now evaluate Eq. (\ref{eq:herm2}) term by term.  The Hamiltonian,
${\hat H}_{QQ}$,
is Hermitian and its eigenvalues are real so we immediately have
${\langle}{\phi}_j|{\hat H}_{QQ}|{\phi}_j{\rangle}
-{\langle}{\phi}_j|{\hat
H}_{QQ}|{\phi}_j{\rangle}^*={\lambda}_j-{\lambda}_j^*=0$. We will assume
boundary conditions
${d{\phi}_j\over dx}{\big|}_{x=a}=0$ and use Eqs. (\ref{eq:surf1}) and
(\ref{eq:surf2}).
     Then
{\it for these special boundary conditions} we find
${\langle}{\Phi}_n|{\hat H}_{PQ}|{\phi}_j{\rangle}=0$ and
${\langle}{\Phi}_n|{\hat H}_{PQ}|{\phi}_j{\rangle}^*=0$. Thus, the
Hermiticity condition
reduces to
\begin{eqnarray}
{\langle}{\Psi}_{\beta}|{\hat H}|{\Psi}_{\alpha}{\rangle}
-{\langle}{\Psi}_{\alpha}|{\hat H}|{\Psi}_{\beta}{\rangle}^* \nonumber\\
={\sum_{j=1}^M}{\sum_{n=1}^N}[{\langle}{\phi}_j|{\hat
H}_{QP}|{\Phi}_n{\rangle}a_j^*B_n
-{\langle}{\phi}_j|{\hat H}_{QP}|{\Phi}_n{\rangle}^*b_jA_n^*] \nonumber\\
+{\sum_{n=1}^N}[{\langle}{\Phi}_n|{\hat H}_{PP}|{\Phi}_n{\rangle}
-{\langle}{\Phi}_n|{\hat H}_{PP}|{\Phi}_n{\rangle}^*]A_n^*B_n=0.
\label{eq:herm3}
     \end{eqnarray}

It is useful now to perform the spatial integrations implicit in Eq.
(\ref{eq:herm3}).
Let us first consider the last term. We can write
${\Phi}_n(x,y){\equiv}{\langle}x,y|{\Phi}_n{\rangle}=
{\chi}_n(x){\sqrt{2\over d+a}}{\sin}{\bigl(}{n{\pi}y\over d+a}{\bigr)}$. If
we substitute into
the last term in Eq. (\ref{eq:herm3}) and perform the integration over $y$,
we obtain
\begin{eqnarray}
{\langle}{\Phi}_n|{\hat H}_{PP}|{\Phi}_n{\rangle}
-{\langle}{\Phi}_n|{\hat H}_{PP}|{\Phi}_n{\rangle}^* \nonumber\\
={-{\hbar}^2\over 2m}{\int_{-\infty}^{0}}dx~
{\biggl[}{\chi}_n^*(x){d^2\over dx^2}{\chi}_n(x)-
{\chi}_n(x){d^2\over dx^2}{\chi}^*_n(x){\biggr]} \nonumber\\
={-{\hbar}^2\over 2m}
{\biggl[}{\chi}_n^*(x){d\over dx}{\chi}_n(x)-
{\chi}_n(x){d\over dx}{\chi}^*_n(x){\biggr]}_{-\infty}^{0}.
\end{eqnarray}
Note also that
\begin{eqnarray}
{\langle}{\phi}_j|{\hat H}_{QP}|{\Phi}_n{\rangle}=C{{\hbar}\over i}
{\int_0^L}dx_1~{\int_0^{{g}(x_1)}}dy_1~{\int_{-\infty}^{0}}dx_0~{\int_0^{d+a}}
dy_0
\nonumber \\
{\times}{\langle}{\phi}_j|x_1,y_1{\rangle}{\delta}(x_1){\delta}(x_1-x_0)
{\delta}(y_1-y_0)
{d\over dx_0}{\langle}x_0,y_0|{\Phi}_n{\rangle}. \label{eq:herm4.0}
\end{eqnarray}
Perform the integration over $x_1$ and notice that ${g}(0)=d+a$.
Then Eq. (\ref{eq:herm4.0}) takes the form
\begin{eqnarray}
{\langle}{\phi}_j|{\hat H}_{QP}|{\Phi}_n{\rangle}=C{{\hbar}\over 2i}
{\int_0^{d+a}}dy_1~{\int_{-\infty}^{0}}dx_0~{\int_0^{d+a}}dy_0
\nonumber \\
{\times}{\langle}{\phi}_j|0,y_1{\rangle}{\delta}(x_0){\delta}(y_1-y_0)
{d\over dx_0}{\langle}x_0,y_0|{\Phi}_n{\rangle}. \label{eq:herm4}
\end{eqnarray}
The cavity basis states, {\it at the interface}, can be written
\begin{equation}
{\langle}{\phi}_j|0,y{\rangle}=
{\phi}^*_{j,n}(0){\sqrt{2\over d+a}}{\sin}{\biggl(}{n{\pi}y\over d+a}{\bigg)}.
\label{eq:eneig5}
\end{equation}
Thus if we perform the
remaining integrations in Eq. (\ref{eq:herm4}), we finally obtain
\begin{equation}
{\langle}{\phi}_j|{\hat H}_{QP}|{\Phi}_n{\rangle}
=C{{\hbar}\over 4i}{\phi}_{j,n}^*(0){d{\rm {\Phi}}_n\over dx}{\biggr|}_0.
\label{eq:herm5}
\end{equation}
We can now combine the above results and write the Hermiticity condition in
the form
\begin{eqnarray}
C{{\hbar}\over 4i}{\sum_{j=1}^M}{\sum_{n=1}^N}
{\biggl(}{\phi}_{j,n}^*(0){d{\chi}_n\over dx}{\biggr|}_0~a_j^*B_n-
{\phi}_{j,n}(0){d{\chi}^*_n\over dx}{\biggr|}_0b_jA_n^*{\biggr)} \nonumber\\
+{-{\hbar}^2\over 2m}{\sum_{n=1}^N}
{\biggl[}{\chi}_n^*(x){d\over dx}{\chi}_n(x)-
{\chi}_n(x){d\over dx}{\chi}^*_n(x){\biggr]}_{-\infty}^{0}A_n^*B_n=0
\end{eqnarray}
{\it for the case of zero-slope boundary conditions for the cavity basis
states}.
The boundary conditions at $x=-{\infty}$ cannot depend on details of the
interface at
$x=0$. Therefore we must satisfy the conditions,
\begin{eqnarray}
C{{\hbar}\over 4i}{\sum_{j=1}^M}{\sum_{n=1}^N}
{\biggl(}{\phi}_{j,n}^*(0){d{\chi}_n\over dx}{\biggr|}_0a_j^*B_n-
{\phi}_{j,n}(0){d{\chi}^*_n\over dx}{\biggr|}_0b_jA_n^*{\biggr)} \nonumber\\
+{-{\hbar}^2\over 2m}{\sum_{n=1}^N}
{\biggl[}{\chi}_n^*(x){d\over dx}{\chi}_n(x)-
{\chi}_n(x){d\over dx}{\chi}^*_n(x){\biggr]}_{x=0}A_n^*B_n=0 \label{eq:bound1}
\end{eqnarray}
and
\begin{equation}
{-{\hbar}^2\over 2m}{\sum_{n=1}^N}
{\biggl[}{\chi}_n^*(x){d\over dx}{\chi}_n(x)-
{\chi}_n(x){d\over dx}{\chi}^*_n(x){\biggr]}_{x=-{\infty}}A_n^*B_n=0,
\label{eq:bound2}
\end{equation}
separately.
The Hermiticity condition, Eq. (\ref{eq:bound1}), is very simply satisfied
if we let
\begin{equation}
C{{\hbar}\over 4i}{\sum_{j=1}^M}{\phi}_{j,n}^*(0)a_j^*
={{\hbar}^2\over 2m}{\chi}_n^*(0)A_n^*
\label{eq:herm7}
\end{equation}
and
\begin{equation}
C{{\hbar}\over 4i}{\sum_{j=1}^M}{\phi}_{j,n}(0)b_j={{\hbar}^2\over
2m}{\chi}_n(0)B_n
\label{eq:herm8}
\end{equation}
These relations will be useful in the next section.

We now have enough information that we can determine the value of
the coupling constant, $C$.  Any state in the waveguide must satisfy the
condition
that it be a continuous function of $x$ and $y$, and that it's slope be a
continuous
function of $x$ and $y$.  Let us consider the state,
$|\Psi_{\beta}{\rangle}$. We require that
\begin{equation}
{\langle}0,y|{\hat Q}|\Psi_{\beta}{\rangle}={\langle}0,y|{\hat
P}|\Psi_{\beta}{\rangle}.
\label{eq:cont1}
\end{equation}
This, in turn, implies that
\begin{equation}
{\sum_{j=1}^M}{\phi}_{j,n}(0)b_j={\chi}_n(0)B_n. \label{eq:cont2}
\end{equation}
If we now compare Eqs. (\ref{eq:herm8}) and (\ref{eq:cont2}), we find that
the coupling constant is given by
\begin{equation}
C={4{\hbar}i\over 2m}. \label{eq:coupconst}
\end{equation}
In Section (4), we use these results to make contact with scattering theory.

\pagebreak

\newpage

\begin{figure}
\caption{The geometry of the two dimensional electron wave guide
used in our calculations.
The rippled waveguide is the region defined with solid lines, rectangular
waveguide is the region whose upper boundary is given by the dotted line. The
dotted-dashed line shows the interference between leads and scattering
region. Here  `a' is the width of the  ripple,
`d' is the width of the rectangular waveguide, scattering cavity
extends from $x=0$
to $x=L$}
\label{figure1}
\end{figure}

\begin{figure}
\caption{ Error in eigenphase, $\theta$,  versus energy for an
R-Matrix
with $N=1000$ and $N=10000$ terms kept in the sum and the FEM calculation of
$\theta$.
The errors are calculated in terms of the fractional deviation of the
numerically computed
eigenphase, ${\theta}_c$, from the exact eigenphase $\theta_e = 2 kL$.}
\label{figure2}
\end{figure}

\begin{figure}
\caption{(a) Wigner delay time, ${\tau}_1$,  for rippled waveguide with
parameters,
$a=10\AA$,
$d=100\AA$, $L=200\AA$, for the energies such that only one
propagating
mode exists in the lead.
    The dashed line is for
reaction matrix calculations,
the solid line is from finite element calculations. (b) Partial Wigner delay
times, ${\tau}_1$ and ${\tau}_2$, when there are two propagating modes in
the leads.
The dashed lines are the reaction matrix
results  and the solid lines are the finite element results.
}
\label{figure3}
\end{figure}

\begin{figure}
\caption{Poles of the S-matrix in the lower complex energy plane
for an energy interval with only one propagating mode in the lead, obtained
    from Eq. (\ref{eq:scat8}). The waveguide parameters are
$a=10\AA$,
$d=100\AA$, $L=200\AA$.}
\label{figure4}
\end{figure}

\begin{figure}
\caption{The effect of evanescent modes for a waveguide with parameters
$a=10\AA$,
$d=100\AA$, $L=500\AA$. (a) The solid
line is Wigner delay time,
$\tau_1$, in the one channel regime, $E_1<E<4E_1$,  for an energy interval
just before a second propagating channel  opens in the lead. The dashed line
shows $\tau_1$ when no evanescent
    modes are included in calculations. (b) The amplitudes, $c_1$ and
$c_2$, of
the first two evanescent
modes in the same energy interval as in (a). (c) The solid line shows the
first
partial Wigner delay time, ${\tau}_1$  just before the opening of the third
channel.
The dashed
line is for the case when no evanescent modes are included in calculating
$\tau_1$. (d)
The same as (c)  for the second partial Wigner delay time, ${\tau}_2$.
}
\label{figure5}
\end{figure}

\begin{figure}
\caption{ The absolute value of S-matrix elements, $|S_{11}|$ and
$|S_{12}|$, in  two modes energy interval, $4E_1<E<9E_1$, where two
propagating channels are allowed.  The waveguide
parameters are $a=10\AA$,
$d=100\AA$, $L=500\AA$.
}\label{figure6}
\end{figure}
\begin{figure}
\caption{ Wigner delay times for the energy interval, $4E_1<E<9E_1$, where two
propagating channels are allowed.
The waveguide
parameters are $a=10\AA$,
$d=100\AA$, $L=500\AA$.
}\label{figure7}
\end{figure}

\begin{figure}
\caption{Poles of $|S_{11}|$ in the lower complex energy plane
for the energy interval, $4E_1<{\rm Re}(E)<9E_1$.
The waveguide
parameters are $a=10\AA$,
$d=100\AA$, $L=500\AA$.  (a) $|S_{11}|$
 for $0<{\rm Im}(E)<-0.95E_1$.
(Only points for which $|S_{11}|<200$ are shown.)
(b) $|S_{11}|$
 for $0<{\rm Im}(E)<-0.075E_1$.}
\label{figure8}
\end{figure}
\begin{figure}
\caption{ The difference between S-matrix element, $|S_{11}|$ with
evanescant modes, and S-matrix element, $|S_{11}^0|$ without
evanescant modes.
The waveguide
parameters are $a=10\AA$,
$d=100\AA$, $L=500\AA$.
}
\label{figure9}
\end{figure}

\begin{figure}
\caption{ The  value of $|S_{11}|$,
when we include the energy dependence of the coupling
matrices ${\bar w}_{Np}$ and ${\bar w}_{Ne}$ (solid line), and its value,
$|S_{11}^{noen}|$,
when we fix the energy dependence of the coupling
matrices ${\bar w}_{Np}$ and ${\bar w}_{Ne}$ to be $E=6.5E_1$ (dotted-dashed
line).  The cavity parameters
The waveguide
parameters are $a=10\AA$,
$d=100\AA$, $L=500\AA$.
}
\label{figure10}
\end{figure}

\begin{figure}
\caption{ The value of $|S_{11}|$, in the complex energy plane, when we
take into account the energy dependence of the coupling matrices ${\bar
w}_{Np}$ and ${\bar w}_{Ne}$ (solid line), and its value,
$|S_{11}^{noen}|$,
when we fix the energy dependence of the coupling
matrices ${\bar w}_{Np}$ and ${\bar w}_{Ne}$ to be $E=6.5E_1$ (dotted-dashed
line).  The cavity parameters
$a=10\AA$,
$d=100\AA$, $L=500\AA$. (Only points for which $|S_{11}|<200$ are shown.)
}
\label{figure11}
\end{figure}
\begin{figure}
\caption{Histograms of Wigner delay times for different numbers
of propagating channels, $M$ for the case of deterministic scattering
((a)-(c))
and RRMT predictions, (d)-(i). (a) Deterministic scattering with $M=2$ for
$a=10\AA$,
$d=100\AA$, $L=500\AA$. (b) Same as (a) for $M=6$.
(c) Same as (a) for $M=16$.
(d) $M=2$ RRMT result with coupling constant, $g=1.8$.
(e) Same as (d) for $M=6$.
(f) Same as (d) for $M=16$.
(g) $M=2$ RRMT result with a unit coupling constant $g=1$.
(h) Same as (g) for $M=6$.
(i) Same as (g) for $M=16$.
}
\label{figure12}
\end{figure}
\end{document}